\documentclass[reprint, amsmath,amssymb, aps, superscriptaddress, floatfix]{revtex4-1}

\usepackage[utf8]{inputenc}
\usepackage{graphicx}
\usepackage{dcolumn}
\usepackage{bm}
\usepackage{braket}
\usepackage{color}
\usepackage{units}
\usepackage{dsfont}
\usepackage{float}
\usepackage{hyperref}
\usepackage{xcolor}
\usepackage{comment}

\begin{document}
\title{Broadband spectral manipulation of single photons using cross-phase modulation}

\author{Kate L. Fenwick}
\affiliation{National Research Council of Canada, 100 Sussex Drive, Ottawa, Ontario K1A 0R6, Canada}
\author{Fr\'ed\'eric Bouchard}
\email{frederic.bouchard@nrc-cnrc.gc.ca}
\affiliation{National Research Council of Canada, 100 Sussex Drive, Ottawa, Ontario K1A 0R6, Canada}
\author{Guillaume Thekkadath}
\affiliation{National Research Council of Canada, 100 Sussex Drive, Ottawa, Ontario K1A 0R6, Canada}
\author{Philip J. Bustard}
\affiliation{National Research Council of Canada, 100 Sussex Drive, Ottawa, Ontario K1A 0R6, Canada}
\author{Duncan England}
\affiliation{National Research Council of Canada, 100 Sussex Drive, Ottawa, Ontario K1A 0R6, Canada}
\author{Benjamin J. Sussman}
\affiliation{National Research Council of Canada, 100 Sussex Drive, Ottawa, Ontario K1A 0R6, Canada}
\affiliation{Department of Physics, University of Ottawa, Advanced Research Complex, 25 Templeton Street, Ottawa ON Canada, K1N 6N5}

\begin{abstract}
Manipulating the frequency and bandwidth of light is crucial in classical and quantum applications including communication, spectroscopy, imaging, and signal processing. Such capabilities also offer potential for interfacing disparate quantum systems in quantum networking and for quantum information processing. We experimentally demonstrate deterministic, broadband frequency control of heralded telecom-band single photons via cross-phase modulation in a short length of single-mode fiber. An intense, ultrafast pump pulse imposes a transient, intensity-dependent refractive-index gradient which imparts a tunable phase shift on the single photons. We present absolute frequency shifts of up to $+6.46\pm0.01$\,THz and $-5.74\pm0.01$\,THz, and bandwidth manipulation ranging from a factor of $0.66\pm0.03$ to $8.4\pm0.3$ times that of the input. Spectral measurements are acquired with a time-of-flight spectrometer and superconducting nanowire detectors.  Our scheme offers a compact and scalable route to spectral routing and bandwidth engineering for ultrafast quantum networking and quantum information processing.
\end{abstract}

\maketitle

Photons occupy a central role across quantum technologies. They are one of the leading candidates for scalable, room-temperature quantum computing architectures~\cite{aghaee2025scaling,alexander2404manufacturable,larsen2025integrated}, the only practical carriers of quantum information over metropolitan and inter-city networks~\cite{chen2021integrated}, and the probes of choice in a wide range of quantum sensing schemes~\cite{abbott2016observation}.  Progress in quantum photonics therefore hinges on high-performance components that can generate, coherently manipulate, and measure quantum states of light. Among the available degrees of freedom, the optical spectrum is of paramount importance~\cite{lu2023frequency}.  It determines loss in transmission media, as well as compatibility with quantum memories and other quantum systems~\cite{maring2017photonic}.  Operating in the telecom C-band simultaneously minimizes fiber attenuation and unlocks state-of-the-art superconducting detectors such as transition-edge sensors and superconducting nanowire single photon detectors (SNSPDs) that provide photon-number resolution and few-picosecond timing jitter~\cite{gerrits2016superconducting,korzh2020demonstration}. Consequently, the ability to translate single photons to telecom wavelengths, or to tailor their bandwidth to match disparate devices, is a key enabler for large-scale photonic quantum networks.

Realizing such spectral control at the single-photon level is challenging. {It has been shown that diamond quantum memories can be used for spectral manipulation of single photons~\cite{fisher2016frequency,bustard2017quantum}, but these demonstrations were limited to the visible regime.} Electro-optic modulators allow for operation bandwidths of up to hundreds of GHz~\cite{wright2017spectral,karpinski2017bandwidth,mittal2017temporal,hiemstra2020pure,hu2021chip,zhu2022spectral,sosnicki2023interface,renaud2023sub,couture2025terahertz}, {while all-optical operation is necessary to push beyond this.} {The low intensity of quantum states of light makes direct second- and third-order all-optical frequency conversion extremely inefficient. Instead, bright ``escort'' pulses are used to drive highly efficient frequency conversion of quantum light. In materials with second-order optical nonlinearities, a single pulse is used to up-convert single photons by sum-frequency generation~\cite{kumar1990quantum,eckstein2011quantum,donohue2014ultrafast,lavoie2013spectral}. In materials with a strong third-order optical nonlinearity, pairs of pulses are used to drive Bragg scattering four-wave mixing~\cite{mckinstrie2005translation,mcguinness2010quantum,tannous2025efficient}. In each case, tailored (quasi-) phase matching is required to achieve high efficiency at a particular frequency. An alternative approach, based on cross-phase modulation (XPM), offers a workaround to this.} In XPM, a bright, ultrashort pump pulse transiently modifies the refractive index experienced by a weak signal photon~\cite{england2021perspectives,kupchak2019terahertz,hall2011ultrafast,fenwick2025ultrafast}. With pump durations in the femto- to picosecond range, XPM enables tunability over several terahertz, more than sufficient to bridge typical mismatches between quantum emitters, memories, and telecom infrastructure. In this letter we exploit broadband XPM to demonstrate deterministic frequency translation and bandwidth manipulation of single photons, providing a versatile, fiber-compatible interface for ultrafast quantum photonic systems.

\begin{figure*}[t!]
	\centering
		\includegraphics[width=0.9\textwidth]{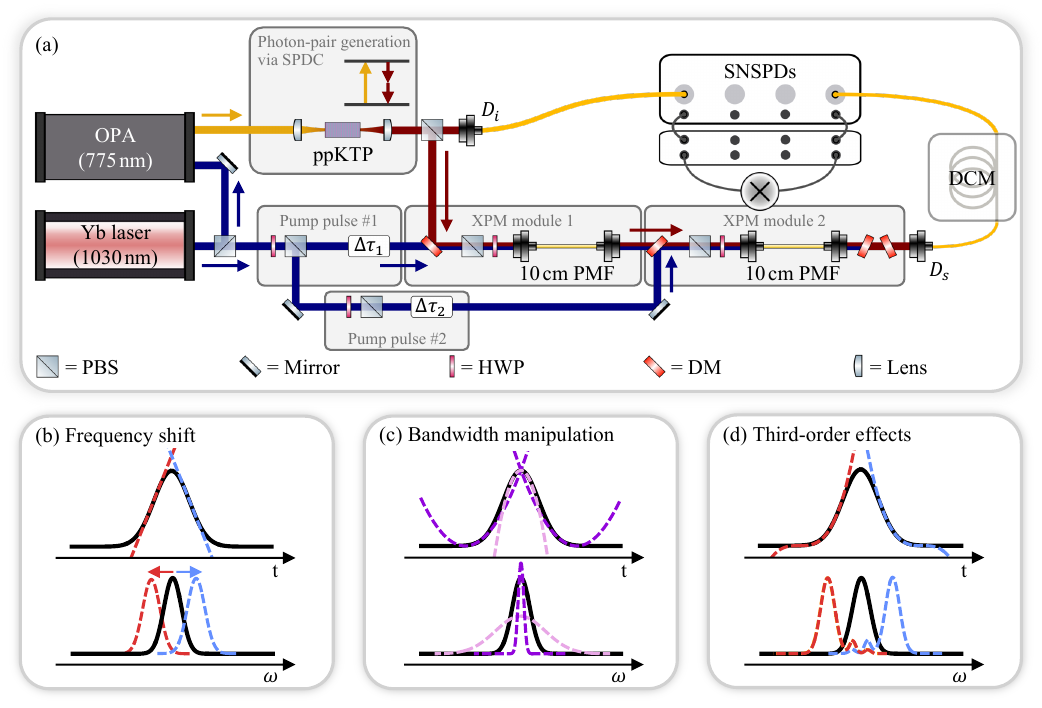}
	\caption{\textbf{Experimental schematic.} (a) A pulsed Ytterbium (Yb) laser pumps an optical parametric amplifier (OPA) and also provides the pump pulses necessary for spectral manipulation via cross-phase modulation (XPM). Output pulses from the OPA pump a periodically poled potassium titanyl phosphate (ppKTP) crystal to generate photon pairs at 1550\,nm via spontaneous parametric down-conversion (SPDC). The idler photons, measured at detector $D_i$, herald the presence of signal photons at detector $D_s$. XPM induced by pump pulses 1 and 2 in short segments of polarization-maintaining single-mode fiber (PMF) is used to manipulate the spectrum of the signal photon, which is measured by a time-of-flight spectrometer configuration. The timing of the signal photon relative to each pump pulse will determine how its spectrum is manipulated: (b) a linear temporal phase will induce a frequency shift; (c) a quadratic temporal phase will modify the bandwidth; and (d) other higher-order effects are possible. PBS: polarizing beamsplitter; HWP: half-wave plate; DM: dichroic mirror; SNSPDs: superconducting nanowire single-photon detectors; DCM: dispersion-compensating module; $\Delta\tau_{1}$, $\Delta\tau_{2}$: pump 1 and 2 delay lines, respectively.}
	\label{fig:expSetup}
\end{figure*}

A key advantage of our approach is its inherently low insertion loss. Since the spectral modulation is performed entirely within standard polarization-maintaining single-mode fiber (SMF), and many quantum photonic systems are already fiber-integrated, coupling losses can be minimized. In particular, some photon-pair sources based on spontaneous four-wave mixing are natively implemented in SMF~\cite{cohen2009tailored}, while other source types, such as those based on nonlinear crystals or integrated platforms, ultimately require fiber coupling to interface with downstream components, including detectors. {Another advantage to operating in standard commercial SMF is the zero-dispersion wavelength, often found between 1300--1400\,nm, as it is straightforward to find a pump pulse that is group velocity matched with the telecom C-band.} The fiber-based geometry of our spectral modulator also lends itself to cascaded modular configurations, providing greater control over the output spectral mode. The temporal intensity profile of the pump pulse (typically Gaussian) can naturally imprint both linear and quadratic temporal phase gradients on the single-photon signal. The sign and strength of these gradients are controlled by the pump pulse energy and the relative pump-signal delay; however, it can be challenging to isolate a purely linear or purely quadratic phase profile due to the fixed shape of the pump. By cascading two XPM modules, residual phase terms from the first module can be compensated for in the second. With four accessible tuning parameters, {\textit{i.e.,}} two pump delays and two pump pulse energies, this scheme enables fine control over both frequency shift and bandwidth shaping, all while maintaining low overall insertion loss through efficient SMF-to-SMF coupling. We note that employing a reconfigurable pulse shaper on the pump would allow its envelope to be tailored, selectively enhancing the linear or quadratic component of its temporal profile.

\begin{figure*}[t!]
	\centering
		\includegraphics[width=0.9\textwidth]{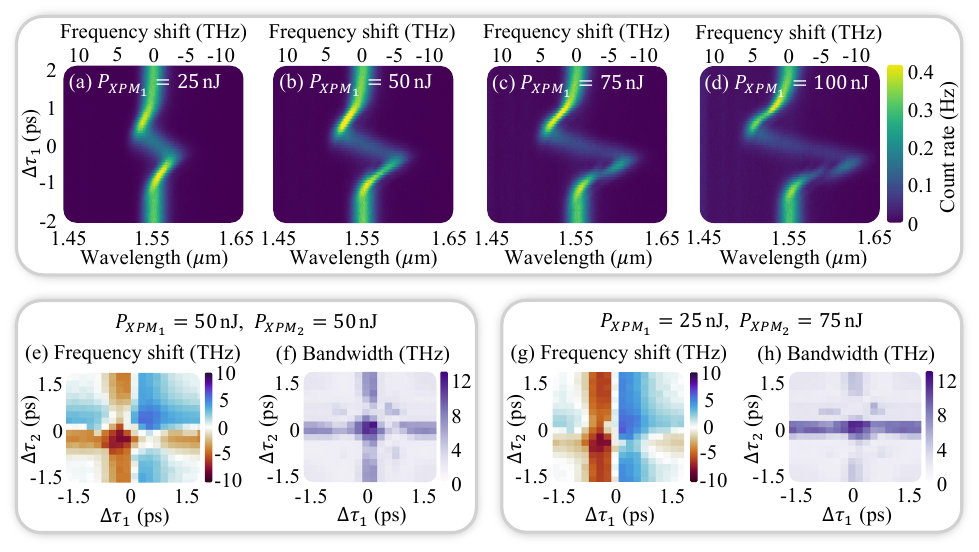}
	\caption{\textbf{Spectral manipulation via XPM.} (a)--(d) Signal photon spectra as a function of pump pulse delay in the first XPM module for increasing pump pulse energy, $P_{\text{XPM}_1}=25$, 50, 75, and 100\,nJ, respectively. (e)--(f) Frequency shift and bandwidth as a function of pump pulse delay in the first ($\Delta\tau_1$) and second ($\Delta\tau_2$) XPM modules, respectively, of the signal photon spectrum when two XPM modules are employed with equal pump pulse energy, $P_{\text{XPM}_1}=P_{\text{XPM}_2}=50$\,nJ. (g)--(h) Frequency shift and bandwidth, respectively, of the signal photon spectrum when two XPM modules are employed with differing pump pulse energies, with $P_{\text{XPM}_1}=25$\,nJ and $P_{\text{XPM}_2}=75$\,nJ.}
	\label{fig:snakePlots}
\end{figure*}

The experimental setup is illustrated in Fig.~\ref{fig:expSetup}(a). A commercial Yb-doped ultrafast laser system (Carbide, LightConversion) produces 180\,fs pulses centered at 1030\,nm with a 200\,kHz repetition rate. A portion of the output is directed to an optical parametric amplifier (OPA, Orpheus-HP, LightConversion), which is tuned to generate 775\,nm pulses that subsequently pump a 2\,mm periodically poled potassium titanyl phosphate (ppKTP) crystal phase-matched for type-II degenerate spontaneous parametric down-conversion (SPDC). This process generates pairs of photons centered at 1550\,nm with orthogonal polarizations. A polarizing beam splitter (PBS) separates the signal and idler photons, and detection of the idler beam serves to herald the presence of a signal photon. The idler photon is detected by an SNSPD system (Single Quantum) with a timing jitter of $\sim20$\,ps, providing us with photon-number resolving capabilities~\cite{cahall2017multi}. We herald only on single-photon idler events, filtering out higher photon-number events. Prior to detection, the signal photons are coupled, in series, into two short segments of standard polarization-maintaining single-mode fiber (PMF, Corning PM1550) where they are made to overlap spatially and temporally with bright pump pulses at 1030\,nm from the driving laser source. Variable delay stages control the relative timing between the signal photon and pump pulses, enabling tunable spectral manipulation via XPM. {The insertion loss of the first XPM module is approximately 2.27\,dB, while only about 0.62\,dB for the second XPM module.} After exiting the two segments of PMF, the signal photons are directed into a time-of-flight spectrometer consisting of a dispersion compensating module (DCM), with dispersion $D=1.033$\,ns/nm, and are then detected by a broadband SNSPD system (Photon Spot) with a timing jitter of $\sim100$\,ps. We use this second, slower SNSPD system for the signal detection as it spans a larger spectral range (1400\,nm to 1650\,nm) than the aforementioned lower jitter SNSPD system used for idler photon detection. 

During each stage of XPM, the timing of the signal photon relative to the temporal profile of the pump pulse will determine how its frequency is manipulated. Based on the Fourier shift theorem, a linear temporal phase gradient produces a frequency shift, while a quadratic phase yields bandwidth compression or expansion (given appropriate chirp on the signal), shown in Figs.~\ref{fig:expSetup}(b) and (c), respectively. Higher-order modulations are also possible---see \textit{e.g.,} the third-order effects depicted in Fig.~\ref{fig:expSetup}(d). Each of these effects can be seen in Fig.~\ref{fig:snakePlots}. We first operate with a single XPM module, where the signal photon spectrum is plotted as a function of pump pulse delay ($\Delta\tau_1$) in Figs.~\ref{fig:snakePlots}(a)--(d) for increasing pump pulse energy, $P_{\text{XPM}_1}$. The amount of spectral shift (both to the blue and to the red) and bandwidth manipulation increases with pump pulse energy. A spectral shift is induced when the signal overlaps with the rising or falling edge of the pump pulse transient ($\Delta\tau_1\sim\pm0.75$\,ps), while bandwidth expansion and compression can be observed near the center of the pump pulse ($\Delta\tau_1\sim0$\,ps) and its tails ($\Delta\tau_1\sim\pm1$\,ps), respectively. Evidence of third-order phase shifts can also be seen, most notably for $P_{\text{XPM}_1}=100$\,nJ and $\Delta\tau_1\sim-0.75$\,ps. {For the pulse durations and fiber lengths used in this work, group velocity walkoff is negligible. This is possible because we operate in a regime in which the pump and signal wavelengths are on approximately opposite sides of the zero-dispersion wavelength.}

Adding a second XPM module can be used to effectively ``double'' the phase shifts achievable with only a single XPM module, as can be seen in Figs.~\ref{fig:snakePlots}(e)--(f). This strategy can be preferable to simply doubling the pump pulse energy, which would amplify parasitic nonlinear noise and self-phase-modulation–induced distortions, leading to unwanted higher-order modulation. Furthermore, with full control over the two XPM modules, the total phase shift experienced by the signal photon can be precisely adjusted. This opens up the ability to use the second XPM module to compensate for any undesired modulation induced by the first XPM module (or vice versa). As shown in Figs.~\ref{fig:snakePlots}(g)--(h), the two XPM modules can be operated with independent pump pulse energies and delays. We exploit this capability to demonstrate frequency shifting while maintaining good spectral overlap with the initial signal photon spectrum. 

\begin{figure}[t!]
	\centering
		\includegraphics[width=0.5\textwidth]{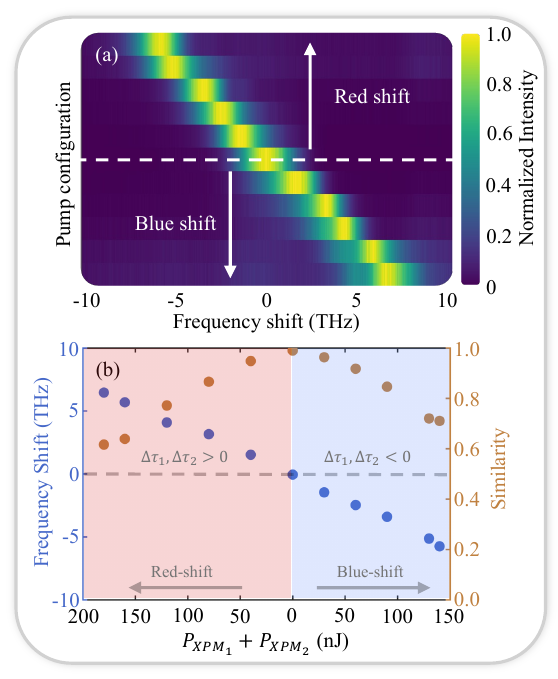}
	\caption{\textbf{Frequency translation.} (a) {Shifted signal spectra for which the pump pulse energy and delay (pump configuration) of each XPM module are tuned to optimize overlap (similarity, $S$) with the input signal spectrum when shifted by the desired amount {(based on pre-selected target shifts in increments of $\pm10$\,nm)}. WhiImportantlyte dashed line corresponds to the initial signal photon spectrum, while white arrows indicate the direction in which total pump pulse energy is increasing.} (b) Shifted central frequency is determined from the maximum value of each spectrum, and is plotted on the left axis (blue), while the similarity is plotted on the right axis (orange). {Note that the total pump pulse energy (x-axis) increases in both directions, relative to the initial signal photon spectrum.} Each spectrum shown in (a) is an average of 20 spectra, each acquired with an integration time of 30\,s. Error bars in (b) are calculated to one standard deviation, and are smaller than the data points used.}
 
	\label{fig:shift}
\end{figure}

Shown in Fig.~\ref{fig:shift}(a), we see both red- and blue-shifted spectra for increasing combined pump pulse energy, $P_{\text{XPM}_1}+P_{\text{XPM}_2}$. The center frequency shift of these spectra are plotted in Fig.~\ref{fig:shift}(b) on the left axis (blue). Note that here the pump pulse energy and delay for each XPM module was optimized to maintain good spectral overlap with the initial signal spectral shape. This spectral overlap is quantified by the similarity, $S$, between the normalized shifted spectrum, $p_i(f)$, and the normalized input spectrum when shifted the desired amount, $q_i(f)$, given by
\begin{equation}
    S = \left( \sum_i \sqrt{p_i(f)~q_i(f)} \right)^2,
\end{equation}
which is plotted in Fig.~\ref{fig:shift}(b) on the right axis (orange). The spectral overlap is maintained above 60\% in all cases. It should be noted that pump noise in the spectral region of interest affects this similarity measurement, especially for high values of combined pump pulse energy. When a subtraction of pump spectral noise is performed, this spectral overlap is maintained above 75\%. The ability to shift the spectrum of a single photon by up to $6.46\pm0.01$\,THz, while maintaining its spectral shape, opens up the possibility of frequency bin encoding and manipulation~\cite{lu2023frequency}.

\begin{figure}[t!]
	\centering
		\includegraphics[width=0.5\textwidth]{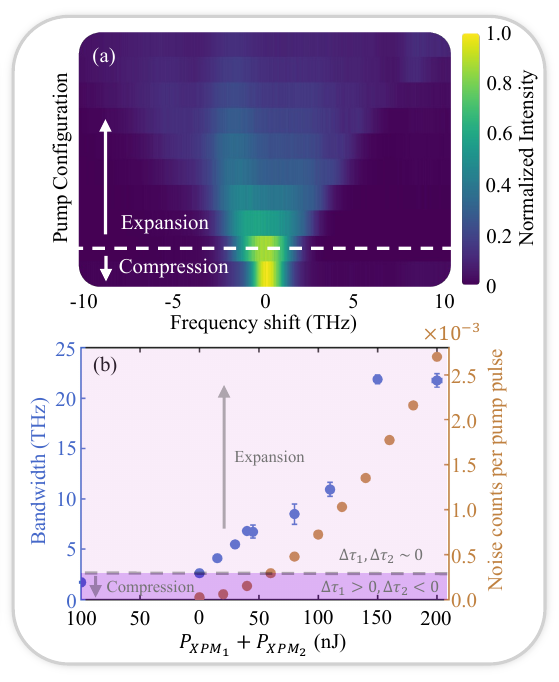}
	\caption{\textbf{Bandwidth manipulation.} (a) {Bandwidth-manipulated signal spectra for which the pump pulse energy and delay (pump configuration) of each XPM module are tuned to optimize for a target bandwidth without inducing a frequency translation. White dashed line corresponds to the initial signal photon spectrum, while white arrows indicate the direction in which total pump pulse energy is increasing.} (b) Bandwidth is determined from the full-width at half-maximum of each spectrum, and is plotted on the left axis (blue). Noise counts per pump pulse are shown on the right axis (orange) for reference. {Note that the total pump pulse energy (x-axis) increases in both directions, relative to the initial signal photon spectrum.} Each spectrum shown in (a) is an average of 20 spectra, each acquired with an integration time of 30\,s. Error bars in (b) are calculated to one standard deviation, and are smaller than some of the data points used.}
 
	\label{fig:bandwidth}
\end{figure}

Beyond this, it is also possible to use our cascaded XPM configuration to achieve both bandwidth expansion and compression of an input signal photon. Note that bandwidth compression is achievable only in cases where the signal has acquired some chirp prior to entering the XPM module. In our experiment, the signal pulse acquires a small positive chirp while traversing the optical components, and this modest chirp is sufficient to induce bandwidth compression when a positive quadratic temporal phase is applied on the Gaussian tail. As shown in Fig.~\ref{fig:bandwidth}(a), we tune the combined pump pulse energy, $P_{\text{XPM}_1}+P_{\text{XPM}_2}$, along with the relative pump pulse delay in each XPM module, in order to manipulate the spectral bandwidth while the central frequency remains unchanged. The spectral bandwidth is plotted in Fig.~\ref{fig:bandwidth}(b) on the left axis (blue), while the noise counts per pump pulse are plotted on the right axis (orange) for reference. 

Importantly, in the case of bandwidth compression, the two XPM modules must be operated such that the signal photon overlaps with the shoulders on opposite sides of the Gaussian peak. This ensures that any linear shift induced by the first XPM module, while compressing the signal spectrum, is reversed by the linear shift induced in the second XPM module. With an initial signal photon bandwidth of $2.6\pm0.1$\,THz, we observe bandwidth compression down to $1.71\pm0.08$\,THz, in agreement with limitations imposed by the time\textit{-}bandwidth product for our signal photons and our chirp estimate. Achieving the observed compression requires $\sim4$\,fs/nm of group-delay dispersion, consistent with the cumulative dispersion of the optical components in our XPM setup. On the other hand, we observe bandwidth expansion up to $21.9\pm0.4$\,THz, which demonstrates the potential of our technique for generating extremely short single photon pulses with broad optical bandwidth. For the observed bandwidth expansion, after re-compression, this could correspond to a single-photon pulse duration below 20\,fs.

We note that the deterministic spectral shift enabled by XPM is equally valuable as a diagnostic tool. In single-photon spectral-shearing interferometry, for instance, imparting a controlled frequency shear between two replicas converts spectral phase into a measurable fringe pattern, allowing full field reconstruction~\cite{davis2018measuring,davis2018experimental,golestani2022electro}. Compared to electro-optic modulation, our XPM-based approach can accommodate significantly larger frequency shears, enabling the characterization of single-photon pulses with much broader bandwidths.

The same XPM-driven shear also functions as an ultrafast, low-loss frequency demultiplexer.  Consecutive sub-picosecond time-bins can be translated to distinct frequency channels with programmable pump delays and then separated using standard wavelength-division components. Such high-efficiency routing is a key primitive for scalable ultrafast photonic quantum circuits~\cite{bouchard2022quantum,bouchard2023measuring,fenwick2024photonic,bouchard2024programmable} that could interleave time- and frequency-bin encodings. Beyond routing, this spectral multiplexing approach can also be leveraged upstream to boost heralding efficiency and state purity in single-photon sources by selecting the indistinguishable frequency mode on demand~\cite{hiemstra2020pure}. Finally, the same broadband, low-loss frequency translation is well-suited to fiber-based quantum memories, where shifting photons into and out of narrow resonant cavities enables efficient write–read cycles~\cite{bustard2022toward}.

Furthermore, many photonic quantum information processing protocols require several phase-synchronized beams at disparate wavelengths. Yet conventional second- and third-order nonlinear frequency conversion processes are acutely sensitive to the pump's optical phase, demanding tight active stabilization. Our scheme, by contrast, is governed solely by pump intensity, so the converted photons acquire a fixed, shot-to-shot phase independent of the pump-carrier phase. This relaxes experimental constraints and simplifies insertion into larger interferometric architectures—including the spectral-shearing schemes mentioned above.

Our results establish broadband XPM as a practical, fiber-compatible means of deterministically translating frequency and manipulating bandwidth at the single-photon level in the telecom band. The demonstrated tunability transforms XPM into a flexible spectral toolbox enabling low-loss routing, demultiplexing, and bandwidth engineering for both classical signals and quantum states, while providing a crucial interface for linking disparate quantum systems in next-generation networks and processors.

\section*{Acknowledgments}
We thank Denis Guay, Khabat Heshami, Aaron Goldberg, Nicolas Couture, Alicia Sit, Nathan Roberts, Noah Lupu-Gladstein, Ramy Tannous, Yingwen Zhang, Milica Banic, Nicolas Dalbec-Constant, Jonathan Baker, Doug Moffatt, and Rune Lausten for their support and insightful discussions.


\providecommand{\noopsort}[1]{}

\end{document}